\documentclass[12pt,pra,aps,amssymb,amsfonts,amsmath,tightenlines]{revtex4}

\usepackage{dsfont} 

\newcommand{\half}{\mbox{$\textstyle \frac{1}{2}$}}
\newcommand{\re}{\mbox{$\rm e$}}
\newcommand{\ri}{\mbox{$\rm i$}}
\newcommand{\rd}{\mbox{$\rm d$}}

\usepackage{xcolor}
\usepackage{bm}
\usepackage{txfonts}
\usepackage{amsfonts}
\usepackage{braket}

\begin{document}

\title{Quantum Measurement of Space-Time Events}
\author{Dorje C. Brody$^{1,2}$ and Lane P. Hughston$^{3}$}
\affiliation{$^1$Department of Mathematics, University of Surrey, 
Guildford GU2 7XH, UK
\vspace{0.1cm}\\
$^2$St Petersburg National Research University of Information 
Technologies, Mechanics and Optics, St Petersburg 197101, Russia
\vspace{0.1cm}\\
$^{3}$Department of Computing, Goldsmiths University of 
London, New Cross, 
London SE14 6NW, UK \vspace{0.2cm}}

\vspace{0.2cm} 
\date{\today}

\begin{abstract}
\noindent 
The phase space of a relativistic system can be identified with the future tube of complexified Minkowski space. As well as a complex structure and a symplectic structure, the future tube, seen as an eight-dimensional real manifold, is endowed with a natural positive-definite Riemannian metric that accommodates the underlying geometry of the indefinite Minkowski space metric, together with its symmetry group. A unitary representation of the 15-parameter group of conformal transformations can then be constructed that acts upon the Hilbert space of square-integrable holomorphic functions on the future tube. These structures are enough to allow one to put forward a quantum theory of phase-space events. In particular, a theory of quantum measurement can be formulated in a relativistic setting, based on the use of positive operator valued measures, for the detection of phase-space events, hence allowing one to assign probabilities to the outcomes of joint space-time and four-momentum measurements in a manifestly covariant framework. This leads to a localization theorem for phase-space events in relativistic quantum theory, determined by the associated Compton wavelength. 
\end{abstract}

\maketitle


\section{Introduction}

\noindent Starting with the pioneering work of Dirac \cite{Dirac}, 
investigations of the Hamiltonian formulation of space-time physics 
have been pursued by numerous authors. One of the motivations behind 
such analysis has been that the mathematical structures of 
phase-space formalisms can be highly amenable to a quantum-mechanical 
description. The naive formulation of a relativistic phase space as a kind of doubled-up 
Minkowski space with four position coordinates and four momentum coordinates, 
while feasible in the classical theory, is not satisfactory as a basis for relativistic quantum theory. Here we propose an alternative approach in which the future tube of  
complexified Minkowski space is taken to be the phase space of a 
relativistic system. Remarkably, this phase space comes naturally equipped with both the symplectic structure and the compatible Riemannian structure needed for the development of a fully covariant relativistic quantum theory. 

Let us write $\mathbb M$ for Minkowski space, by which we mean 
$\mathds R^4$ equipped with the usual flat space-time metric $g_{ab}$ with signature $(+,-,-,-)$. For the positions of points $x, y \in \mathbb M$ relative to an origin in $\mathbb M$ we write $x^a$ and $y^a$, where $a,b = 0, 1, 2, 3$. We say that $x$ and $y$ are time-like, space-like, or null separated according to whether   $g_{ab}(x^a-y^a)(x^b-y^b)$ is positive, negative, or zero. In the time-like and null cases, the separation vector $v^a = x^a - y^a$ is said to be future-pointing or past-pointing according to whether $v^0$ is positive or negative.  Then by complex Minkowski space ${\mathds C}\mathbb{M}$ we mean 
$\mathds C^4$ equipped with the same metric tensor. The so-called future tube ${ \Gamma}$ is the open submanifold of  ${\mathds C}\mathbb{M}$ consisting of points  that are of the form $z^a=x^a-\ri r^a$, where $r^a$ 
is time-like and future pointing. Thus for all $z^a \in {\Gamma}$ it holds that $g_{ab}r^ar^b>0$ with $r^0>0$.
 
The future tube plays an important role in rigorous treatments of quantum 
field theory. In particular, the 
Wightman functions -- given by vacuum expectations of field operators -- are 
analytic in ${\Gamma}$, and  
one can reconstruct the field theory from the data of these 
expectation values \cite{Wightman 1956, Wightman 1960, Araki 1961, 
Streater and Wightman 1964}. The future tube contains no real 
space-time points; however, the so-called extended future tube, consisting of points attainable 
by the action of the complex Lorentz group on ${\Gamma}$, 
contains real points, called Jost points. One can then recover 
the field theory from the values of the Wightman functions at Jost points \cite{Schweber}. Complexified Minkowski space also plays 
an important role in the Penrose twistor program \cite{Penrose 1967, Penrose 1968, Penrose 1972}, as does the future tube. In twistor theory, the complex projective space ${\mathds C}
{\mathbb P}^3$ is divided into two parts, the upper and lower half of ${\mathds C}
{\mathbb P}^3$, separated by a five real dimensional hypersurface $\mathbb N^5$ of null twistors. The points of ${\mathds C}\mathbb{M}$ correspond to complex projective lines in ${\mathds C}
{\mathbb P}^3$. The points of ${ \Gamma}$ correspond to lines that lie entirely in the top half of ${\mathds C}
{\mathbb P}^3$.

In both twistor theory and quantum field theory, the complexification of Minkowski space, natural as it may be, is introduced primarily to enable one to exploit the 
tools of complex analysis in relation to the positive frequency condition on fields; and there is no direct physical significance 
attached as such to the imaginary components of complex
space-time points. Some form of reality condition has to be brought 
into play to make 
the link to the physical ``real'' spacetime. 

From the view of the complex formulation of classical mechanics \cite{Mackey},
 it is natural to ask whether the imaginary part of a  point in 
${\Gamma}$ is related to the four momentum of a relativistic system. In what 
follows we offer an affirmative answer to this question. This, in turn, allows us 
to construct a Hilbert space of quantum states over the space-time phase space. 
The conformal transformations of the underlying space-time phase space can 
then be represented explicitly in terms of unitary operators acting on quantum 
states. We then formulate a measurement postulate for the detection of the 
phase-space location of a relativistic event by identifying the probability law for
measurement outcomes along with an appropriate post-measurement transformation rule for the states. 
It 
is shown, in particular, that when the measurement outcome yields a 
phase-space point, the state results in a coherent state of the conformal group, 
centered at that point. The fact that coherent states are the most localized 
states in the Hilbert space then leads to a localization bound which shows for systems 
of short Compton wavelength
that when an event is detected to have occurred at a specific phase-space 
point, the resulting state will be highly localized in phase space. 

The problem of relativistic quantum measurement has been 
investigated by many authors (see e.g.~\cite{Landau Peierls 1931, Aharonov Albert 1981, Rovelli, Peres} and references cited 
therein). It is often the case, however, that measurement postulates of nonrelativistic quantum theory are used in a relativistic setup to deduce 
implications of the postulates, which is unsatisfactory, for what is required 
is a measurement postulate in a relativistic setup, as we propose here. 

\section{Relativistic mechanics} 

\noindent We begin by reviewing the cotangent bundle approach to 
relativistic mechanics \cite{Todorov,Woodhouse,Marsden,SSG}. 
The phase space is taken to be the cotangent bundle of Minkowski space, where the cotangent vectors in the fibre over a point in Minkowski space are identified with the momentum four-vectors that the particle might possess. The bundle is an eight-dimensional manifold $T^*\mathbb M$, with base coordinates $x^a$ and fibre coordinates $p_a$. We form the canonical one-form 
 $\theta = p_a \rd x^a$ on $T^*\mathbb M$ along with its exterior derivative, the associated symplectic form   
 $\omega = \rd p_a \wedge \rd x^a$. 
 Given a smooth function $H: T^*\mathbb M \to \mathds R$ we then write Hamilton's equations for a dynamical trajectory 
\begin{eqnarray} 
 s \in {\mathds R}^+ \mapsto \{x^a(s), p_a(s)\} \in T^*\mathbb M
 \end{eqnarray} 
 in the form 
\begin{eqnarray} 
\frac{\rd x^a}{\rd s}= \frac{\partial H}{\partial p_a} 
\quad {\rm and} \quad 
\frac{\rd p_a}{\rd s}=- \frac{\partial H}{\partial x^a}\,,  
\label{Hamilton's equations 1} 
\end{eqnarray} 
and we call $H(x^a, p_a)$ the Hamiltonian function. As before, we let $g_{ab}={\rm diag}(+1,-1,-1,-1)$ be the 
metric on the base space ${\mathbb M}$, which we can use to raise and lower indices on the fibre elements as well. Then we can write  $(x^a, p^a) = (x^a, g^{ab}p_b)$ and put Hamilton's equations in the more symmetrical form
\begin{eqnarray} 
\frac{\rd x^a}{\rd s}=g^{ab} \frac{\partial H}{\partial p^b} 
\quad {\rm and} \quad 
\frac{\rd p^a}{\rd s}=-g^{ab} \frac{\partial H}{\partial x^b}\,. 
\label{Hamilton's equations 2} 
\end{eqnarray} 
The cotangent-bundle approach works well 
for characterizing the dynamics of  typical mechanical systems in 
space-time. To convince oneself it suffices to explore a few examples. 

\vspace{0.2cm} 
\noindent {Example 1}: \textit {Free particle.}
The Hamiltonian is taken to be 
\begin{eqnarray} 
H=[g_{ab}p^ap^b]^{1/2}.
\end{eqnarray} 
Then $H$ will be a constant of the motion which we identify as the mass $m$. The phase space is foliated by surfaces of constant $H$, and as an initial condition we choose $\{x^a(0), p_a(0)\}$ to lie on the surface $H = m$. Hamilton's equations (\ref{Hamilton's equations 2}) imply 
\begin{eqnarray} 
m {\dot x}^a=p^a, \quad {\dot p}^a=0.
\end{eqnarray} 
The phase-space trajectory is given by 
\begin{eqnarray}
x^a(s) = x^a(0) + s \,m ^{-1} p_a(0), \quad p_a(s) = p_a(0),
\end{eqnarray} 
 corresponding to a geodesic motion in Minkowski space subject to the specified initial conditions.

\vspace{0.2cm} 
\noindent {Example 2}:  \textit {Charged particle in an electromagnetic field.}
Let the charge be $q$ and write $A^a(x)$ for the electromagnetic four-potential. With the familiar minimal coupling, we extend the previous example by taking the Hamiltonian to be of the form 
\begin{eqnarray} 
H = \left[ g_{ab}( p^a-q A^a)(p^b-q A^b)\right]^{1/2}. 
\end{eqnarray} 
We foliate the phase space with surfaces of constant $H$, identifying the value of $H$ with the mass of the particle. Hamilton's 
equations give 
\begin{eqnarray}
m{\dot x}^a = (p^a-qA^a), \quad
{\dot p}^a = q {\dot x}_c \nabla^a A^c, 
\end{eqnarray} 
where $\nabla_a=\partial/\partial x^a$.  Further differentiation leads to the Lorentz force law 
\begin{eqnarray}
{m \ddot x}^a =  q F^{ab} \, {\dot x}_b, \quad F^{ab} = \nabla^a A^b - \nabla^b A^a.
\end{eqnarray} 

\noindent {Example 3}: \textit {Relativistic two-body problem with a force of mutual attraction.}
Let us write $x^a$, $y^a$, $X^a$, $Y^a$ for the space-time positions and 
momenta of the two particles, setting 
\begin{eqnarray} 
q^a = \half(x^a-y^a), 
\end{eqnarray} 
along with
\begin{eqnarray} 
P^a = X^a + Y^a, \quad
Q^a = X^a - Y^a.
\end{eqnarray} 
To model a central force we project $q^a$ onto the space-like hypersurface orthogonal to the 
total momentum $P^a$ to measure the separation of the two particles. Since $P^a$ is time-like, the resulting ``internal'' coordinate $\xi^a$ defined by
\begin{eqnarray} 
\xi^a = q^a - \frac{q_cP^c}{P_cP^c} \, P^a 
\end{eqnarray} 
is space-like.  Thus $\xi_a\xi^a\leq0$ and 
for the potential we set 
\begin{eqnarray} 
V(\xi^a) = \Phi(-\,\xi_a\xi^a) ,
\end{eqnarray} 
for some
function $\Phi: {\mathds R}^+ \backslash\{0\} \to {\mathds R}$ in $\rm{C}^1(0, \infty)$. For example, for a harmonic oscillator 
we set $\Phi(u)=ku$ where $k \in {\mathds R}^+$. For a Coulomb potential 
set $\Phi(u)= -e_1 e_2 u^{-1/2}$, where $e_1,e_2$ are the charges of 
the particles. For a gravitational potential, set 
$\Phi(u)=Gm_1m_2u^{-1/2}$, and so on. Consider now a pair of particles 
interacting via the potential $V$. We write
\begin{eqnarray} 
X^2 = m_1^{\,2} + V, \quad Y^2 = m_2^{\,2} + V, 
\end{eqnarray} 
where $X^2=X_aX^a$ and $Y^2=Y_aY^a$, and 
$m_1$, $m_2$ are the rest masses. These 
conditions imply
\begin{eqnarray} \half (P^2 + Q^2) - 2V = m_1^{\,2}+m_2^{\,2}, \quad 
P_c Q^c = m_1^{\,2} - m_2^{\,2}. 
\end{eqnarray} 
Hence for the Hamiltonian we set 
\begin{eqnarray} 
H = \left( \half (P^2 + Q^2) - 2V \right)^{1/2} .
\end{eqnarray} 
Since $H$ will be a constant of the motion, we choose the initial 
conditions so that $\{r^a(0), q^a(0)\}$ lies on the surface
\begin{eqnarray} 
H=[m_1^{\,2}+m_2^{\,2}]^{1/2}\,.
\end{eqnarray} 
Hamilton's 
equations show that $P_cQ^c$ is also 
a constant of the motion, so we set 
\begin{eqnarray}
P_c Q^c = m_1^{\,2} - m_2^{\,2}
\,, 
\end{eqnarray} 
thus fixing the two masses. A calculation then shows that 
\begin{eqnarray} 
\frac{\rd^2\xi^a}{\rd s^2} = -\frac{1}{m_1^{\,2}+m_2^{\,2}} \, \Phi' 
(-\xi_c\xi^c) \, \xi^a ,
\end{eqnarray} 
where $\Phi'(u)=\rd\Phi(u)/\rd u$. Since the right side is a function of 
$\xi^a$, we can solve for $\xi^a(s)$, which in turn allows us to 
determine the phase-space trajectory. 
For example, in the case of an oscillator, we have $\Phi'=k$, so we obtain 
\begin{eqnarray} 
\xi^a(s) = \alpha^a \cos(\omega s) + \beta^a \sin(\omega s), 
\end{eqnarray} 
where $\alpha^a$, $\beta^a$ are constant spacelike vectors such that 
$\alpha^a=\xi^a(0)$ and $\omega\beta^a={\dot\xi}^a(0)$, with  
$\omega^2=k/(m_1^{\,2} + m_2^{\,2})$. 
\vspace{0.1cm} 

Despite the merits of these examples, there are limitations to the effectiveness of the cotangent bundle approach as a foundation for the theory of relativistic dynamics. For a start, 
there is no intrinsic mechanism to prevent the momentum from becoming space-like or past-pointing. This problem can be avoided in specific examples, such as the ones above, but it is undesirable that one should have to manage the situation on 
an \textit{ad hoc} basis. The cotangent bundle approach also poses problems when we look at field theories, since 
the cotangent bundle does not admit a natural complex structure. In particular, 
there is no general recipe for combining position and 
momentum in a linear way, allowing one to write complex expressions of the form $x^a+\ri p^a$.
Such variables arise  in the 
quantization of oscillators, but in that case there is a dimensional constant that allows one to modify the expressions to produce terms of the same dimensionality. If the constants of nature at ones disposal are the speed of light and Planck's constant, then one cannot convert a quantity with units of momentum to one with units of length. Further,  the 
interpretation of the parameter $s$ as a proper time in the cotangent bundle approach is ambiguous when many particles are involved. 

What is the best way forward? 
Many authors have considered the problems  arising with relativistic phase spaces, both for classical theories and quantum theories 
\cite{Currie,Kaiser,King,Komar,Rohrlich}. 
Our approach  incorporates ideas drawn from all of these, and from geometric quantum mechanics as well \cite{Kibble 1979, Page 1987, 
Gibbons 1992, Hughston 1995, Ashtekar Schilling 1998, Brody Hughston 2001, Bengtsson Zyczkowski 2006}. We also look closely at the role of probability in the course of our development of a relativistic  theory of quantum measurement based on the geometry of the future tube. 

\section{Back to the future tube} 

\noindent That there is an appropriate map from the cotangent bundle to the future 
tube is not 
immediately apparent, but a dimensional argument will lead the way. In order for us to be able to regard $x^a-\ri r^a$ as a complex phase-space variable in 
a relativistic context we shall require $r^a$ to have units of inverse 
momentum. Then if we multiply $r^a$ by Planck's constant we obtain a vector with units of position that can be combined with $x^a$. 
Specifically, we consider the Kelvin inversion  
\begin{eqnarray}
r^a = \hbar p^a/(p_cp^c), \quad 
p^a=\hbar r^a/(r^ar_a). 
\label{phase-space correspondence}
\end{eqnarray} 
This transformation maps the 
cone of time-like future-pointing Minkowski space vectors 
into itself. Thus we have 
\begin{eqnarray}
\hbar \, g^{ab}\frac{\partial}{\partial p^b} = r_c r^c \left( g^{ab} - 
\frac{2\, r^a r^b}{r_c r^c} \right) \frac{\partial}{\partial r^b} 
\end{eqnarray} 
and 
\begin{eqnarray}
\frac{1}{\hbar} \,g_{ab}\frac{\rd p^a}{\rd s} =  \frac{1}{r_c r^c} \left( g_{ab} - 
\frac{2\, r_a r_b}{r_c r^c} \right) \frac{\rd r^a}{\rd s} .  
\end{eqnarray} 
Now define a symmetric tensor $h_{ab}$ with inverse $k^{ab}$ by setting
\begin{align}
h_{ab} = -\frac{1}{r_c r^c} \left( g_{ab} - \frac{2\, r_a r_b}{r_c r^c} \right), 
\, \, \, \,
k^{ab} = - r_c r^c \left( g^{ab} - \frac{2\, r^a r^b}{r_c r^c} \right) . 
\label{phase-space metric} 
\end{align} 
%
Then $k^{ab}\,h_{bc}=\delta^a_{c}$, and a straightforward calculation shows that Hamilton's equations on the future tube
take the form 
\begin{eqnarray} 
\hbar \, \frac{\rd x^a}{\rd s}= - k^{ab} \frac{\partial H}{\partial r^b} 
\quad {\rm and} \quad 
\hbar \, \frac{\rd r^a}{\rd s}= k^{ab} \frac{\partial H}{\partial x^b} . 
\label{Hamilton's equations on the future tube} 
\end{eqnarray} 
That the signs in  \eqref{Hamilton's equations on the future tube} are reversed in comparison with  \eqref{Hamilton's equations 2} is an artefact of the convention that defines the future tube by points of the form $x^a - \ri r^a$ with $r^a$ time-like and future-pointing. 
But what is not so obvious, and comes perhaps as a surprise, is that  the  
quadratic form $h_{ab}$ is \textit{positive definite}, thus defining a 
Riemannian metric on the future tube, given by 
\begin{align}
\rd s^2 = h_{ab} \, (\rd x^a \, \rd x^b + \rd r^a \, \rd r^b).
\end{align} 
As a consequence we see that the arc-length along a smooth curve can be taken as a canonical parametrization of the phase-space trajectory. In particular, in situations where two or more particles are interacting, the phase space of the system as a whole can be taken to be the product of the phase spaces of the individual systems, with an overall positive definite metric, thus leading to a natural way of synchronizing the dynamics of the constituents.  

\section{Relativistic phase-space geometry} 

\noindent As a number of authors have pointed out, there are several distinct 
but ultimately equivalent ways of arriving at the geometrical structure of the 
future tube  
\cite{Uhlmann1, KN, Uhlmann3, Ruhl, Ruhl2,Odzijewicz, Odzijewicz_1976, Carey 1977, Coquereau 1990, Vladimirov, Gibbons}. 
Building on these works, we pursue here an alternative approach to the geometry of ${\Gamma}$ that ties in naturally with quantum measurement theory. We begin with the Hilbert 
space ${\mathcal H}={L}^2(\Gamma,{\mathcal O})$ of square-integrable 
holomorphic functions on the future tube. If we let  $f, g $ be elements of 
${\mathcal H}$, then for their inner product we write 
\begin{eqnarray}
\langle  \bar g \,| f \rangle = \int_{\Gamma} \! f(z) \,{\bar g}({\bar z}) \,\rd \mu_z,
\end{eqnarray} 
where 
\begin{eqnarray} 
\rd \mu_z = \frac{1}{16}\,\rd^4 z \, \rd^4 {\bar z}
\end{eqnarray} 
denotes the usual Lebesgue measure on 
$\Gamma$. 
The fact that such functions constitute a Hilbert space is nontrivial, for it is not immediately obvious that any Cauchy sequence in  ${L}^2(\Gamma,{\cal O})$ converges to an element of ${L}^2(\Gamma,{\mathcal O})$. That such convergence holds follows as a consequence of a well-known bound \cite{Krantz}, which states that  for any compact subset $Q\subset \Gamma$ there exists a constant $C_Q$ such that for all $\phi \in {L}^2(\Gamma,{\cal O})$ we have 
\begin{eqnarray}
\sup_{z \in Q} |\phi(z)| \leq  C_Q \, ||\phi||,
\label{inequality}
\end{eqnarray} 
where 
\begin{eqnarray}
||\phi|| = \left( \int_{\Gamma} \! \phi(z) \,{\bar\phi}({\bar z}) \,\rd \mu_z \right)^{1/2}.
\end{eqnarray} 
Now let $\{\phi^n\}_{n \in \mathds N}$ be an 
orthonormal basis for ${\mathcal H}$ so that 
\begin{eqnarray} 
\int_{\Gamma} \! \phi^n(z) \,{\bar\phi}_m({\bar z}) \,\rd \mu_z = 
\delta^n_m . 
\end{eqnarray} 
We introduce the Bergman kernel \cite{Meschkowski,Bergman} 
on $\Gamma$ by setting
\begin{eqnarray} 
K(z,{\bar w}) = \sum_n \, {\bar\phi}_n({\bar w}) \, \phi^n(z)  ,
\label{Bergman kernel0}
\end{eqnarray}
which is independent of the choice of orthonormal basis. Thus $K(z,{\bar w})$ is holomorphic in $z$ and antiholomorphic in $w$, and for 
any holomorphic function $f\in{\cal H}$ we evidently have 
\begin{eqnarray} 
\int_{\Gamma} \! K(z,{\bar w}) \,f(w) \, \rd \mu_w = f(z).
\label{reproducing property}
\end{eqnarray}
We thus see that the Bergman kernel acts as an identity operator or reproducing kernel on ${\cal H}$. In particular, for all $x, y, z \in \Gamma$ we have the identity
\begin{eqnarray}
\int_{y }K(x,{\bar y}) \, K(y,{\bar z}) \, \rd \mu_y = K(x, {\bar z}). 
\label{fundamental identity}
\end{eqnarray}
Now consider a smooth curve 
\begin{eqnarray} 
\gamma: \sigma\in [0, 1] \mapsto w_{\sigma} \in \Gamma. 
\end{eqnarray}
For each value of the parameter $\sigma$ 
the function $\psi_{\sigma}: \Gamma \to {\mathds C}$ defined by 
\begin{eqnarray} 
\psi_\sigma(z) = K(z,{\bar w}_{\sigma}) 
\end{eqnarray}
is holomorphic and square integrable. 
It follows that $\psi_\sigma$ describes a curve in 
${\mathcal H}$ as $\sigma$ varies, so we can work out the 
length along $\gamma$ by use of the Fubini-Study metric \cite{KN, Page 1987} : 
\begin{align} 
\rd s^2 =  \frac{\int \rd{\psi}_\sigma(z) \,
\rd{\bar\psi}_\sigma({\bar z}) \, \rd \mu_z}
{\int {\psi}_\sigma(z) \,
{\bar\psi}_\sigma({\bar z}) \, \rd \mu_z} 
- \frac{\left| \,\int \rd{\psi}_\sigma(z) \,
{\bar\psi}_\sigma({\bar z}) \, \rd \mu_z \,\right|^{\,2}}
{\left( \, \int {\psi}_\sigma(z) \,
{\bar\psi}_\sigma({\bar z}) \, \rd \mu_z \,\right)^{\,2}} \, .
\end{align}  
A calculation then shows that
\begin{eqnarray}
\rd s^2 = \frac{\partial^2  \log K(w, \bar w)} {\partial w^a \, \partial \bar w^b } \,
\rd w^a \, \rd{\bar w}^{b} . 
\label{Bergman metric} 
\end{eqnarray}  
Thus, the Fubini-Study metric on $\mathcal H$ induces a K\"ahler metric on 
$\Gamma$. This is the so-called Bergman metric \cite{Bergman}. The ideas 
of Bergman kernel and the associated metric can be elucidated by considering 
an elementary example of a bounded domain ${\mathds C}^+$ of 
${\mathds C}$ defined by $\ri(z-{\bar z})>0$. Writing $z=x-\ri r$ the domain 
can be expressed as the half plane for which $r>0$. Thus 
${\mathds C}^+$ can be thought of as representing one-dimensional future 
tube, i.e. phase space in zero space dimensions, with only time and energy. 
An example of a set of orthonormal functions on ${\mathds C}^+$ is given by 
\begin{eqnarray} 
\phi^n(z) =  2\ri \sqrt{ \frac{n}{\pi}} \frac{(\ri-z)^{n-1}}{(\ri+z)^{n+1}} ,
\end{eqnarray} 
from which one can easily verify that the corresponding Bergman kernel takes 
the form 
\begin{eqnarray}
K_{{\mathds C}^+} (z,\bar w) =\sum_{n=1}^\infty\phi^n(z) \, {\bar \phi_n(\bar w)} = 
\frac{1}{\pi(z-{\bar w})^2} .
\end{eqnarray} 
The associated Bergman metric is then the usual hyperbolic metric of 
the half plane. Note that under the map $w=f(z)$ given by the Cayley 
transform
\begin{eqnarray}
f(z) = \ri\left( \frac{1-z}{1+z}\right) , 
\label{Cayley} 
\end{eqnarray} 
points of ${\mathds C}^+$ are mapped invertibly to points of the unit disk 
$|w|<1$, which is the so-called Poincar\'e disk ${\mathbb D}$. The 
orthonormal basis elements are then mapped to 
\begin{eqnarray}
\phi_n(w) = \sqrt{\frac{n}{\pi}} \, w^{n-1} ,
\end{eqnarray}
and a short calculation shows that the kernel function becomes 
\begin{eqnarray}
K_{\mathbb D}(z,\bar w) = \frac{1}{\pi(1-z{\bar w})^2}\,  . 
\end{eqnarray}

In the case of the future tube, the Bergman kernel can be worked out 
explicitly, and we have
\begin{eqnarray} 
K(z,{\bar w}) =  \left(\frac{2^3\cdot 4 !}{ \,\pi^4} \right) 
\frac{1}{[g_{ab}(z^a-{\bar w}^a)(z^b-{\bar w}^b)]^4} \, . 
\label{Bergman kernel}
\end{eqnarray} 
Note that similar to case of the half plane ${\mathds C}^+$, the future tube 
can be mapped to a ``unit ball" domain ${\mathbb B}$, which acts as the 
higher-dimensional analogue of the Poincar\'e disk ${\mathbb D}$. 
Specifically, if we write points of $\Gamma$ in the $2\times2$ matrix form 
\begin{eqnarray} 
{\hat Z}=z^0{\hat I}+z^1{\hat\sigma}_x+z^2{\hat\sigma}_y+z^3
{\hat\sigma}_z \, , 
\end{eqnarray} 
where ${\hat I}$ denotes the $2\times2$ identity matrix and 
${\hat\sigma}_x$, ${\hat\sigma}_y$, ${\hat\sigma}_z$ are the Pauli 
matrices, the future tube is defined by the condition that 
\begin{eqnarray} 
{\hat I}-
{\hat Z}^\dagger{\hat Z}>0, 
\end{eqnarray} 
where the inequality here means that the left 
side is a positive-definite matrix. The Cayley transform of the future tube in 
this representation is given by 
\begin{eqnarray} 
W = \ri ({\hat I}-{\hat Z}) \, ({\hat I}+{\hat Z})^{-1} , 
\end{eqnarray} 
which is the analogue of (\ref{Cayley}) for the half plane. 
With this transformation in mind, the numerical factor 
$V_4=\pi^4/(2^3\cdot4!)$ appearing in (\ref{Bergman kernel}) can then be 
seen as arising from the Euclidean volume of the ball domain \cite{Hua}. In 
the literature of the future tube, it is common to adopt the characterization 
in terms of ${\mathbb B}$, since this allows for a somewhat simpler 
treatment of the group-theoretic analysis associated with conformal 
transformations 
\cite{Uhlmann1,KN,Uhlmann3,Odzijewicz_1976,Ruhl,Ruhl2,Carey 1977}. 
However, for in the consideration of 
quantum theory we find it more transparent to work directly with the 
standard characterization of the future tube as the domain 
$\Gamma$ in the complexified Minkowski space. In particular, for its 
Bergman metric, 
substitution of \eqref{Bergman kernel} into \eqref{Bergman metric} gives 
\begin{eqnarray}
\frac{\partial^2  \log K(z, \bar z)} {\partial z^a \, \partial \bar z^b } = h_{ab},
\end{eqnarray}
where $h_{ab}$ turns out to be none other than the metric 
(\ref{phase-space metric}) that we introduced earlier using the Kelvin transformation. Since a Bergman metric is fully determined by the complex analytic structure of the underlying domain, it follows that $h_{ab}$ admits the symmetry group of $\Gamma$, which is the 15-parameter conformal group of Minkowski space. These phase-space symmetries are generated by Hamiltonian flows on $\Gamma$.

\section{Quantum states} 

\noindent Going forward, now let $u, v, w, x, y, z$ denote points of $\Gamma$. Having introduced the Hilbert space ${\cal H}$ of holomorphic functions on
$\Gamma$ we are in a position to 
build a quantum theory. A general state will be a density matrix $\rho(y,{\bar z}) \in 
{L}^2(\Gamma,{\cal O})\times {L}^2(\Gamma,\bar{\cal O})$. For such a state, we require the following: (a) that $\rho(y,{\bar z}) = \bar \rho({\bar z},y)$, (b) that $\rho(y,{\bar z})$ should be 
positive, that is to say
\begin{eqnarray}
\int \!\bar\alpha(\bar y) \, \rho(y,{\bar z}) \, \alpha(z)\, \rd \mu_y \, 
\rd \mu_z \geq 0 
\end{eqnarray}
for $\alpha(z)\in{L}^2(\Gamma,{\cal O})$, and (c) that it should have unit trace, 
\begin{eqnarray}
\int \!K(z,{\bar y}) \, \rho(y,{\bar z}) \, \rd \mu_y  \rd \mu_z = \int \! \rho(z,{\bar z}) \, \rd \mu_z = 1 .
\end{eqnarray}
A state is then said to be 
pure if $\rho(y,{\bar z})
=\xi(y)\,\bar \xi(\bar z)$ for some holomorphic function 
$\xi \in{L}^2(\Gamma,{\cal O})$ with unit norm. We observe that for both pure and mixed states the ``diagonal" function $\rho(z,{\bar z})$ takes the form of a \textit {probability density} on $\Gamma$.  That a probability density 
function on phase space arises naturally in the present context is significant, since the construction of such densities in configuration-space models for relativistic quantum mechanics is known to be problematic. 

The interpretation of a density matrix is that it represents {\em the quantum state of a relativistic event}. 
Such an event  is accompanied by position and momentum data. The fact that 
wave functions are holomorphic then prohibits the 
possibility that they can be concentrated with arbitrarily high precision in a given region of phase space. This follows from the fundamental inequality \eqref{inequality}. Many aspects of the theory can be understood as being analogous to  the Bargmann-Segal construction in nonrelativistic quantum mechanics \cite{Bargmann, Segal}. 

\section{Space-time transformations} 

\noindent It is natural to enquire how space-time transformations are represented in ${\mathcal H}$. That is, we are interested in constructing a unitary representation of the action of
the Poincar\'e group $\mathcal P (\Gamma)$ on $\Gamma$.  Such transformations are of the form
\begin{eqnarray} 
P: z^a \mapsto L^a_b\, z^b + B^a 
\label{translation}
\end{eqnarray}  
where $B^a$ is a real four vector and $L^a_b$ satisfies 
\begin{eqnarray} 
g_{ab}L^a_c L^b_d=g_{cd}\, .
\label{Lorentz transformation}
\end{eqnarray}  
 A family of unitary 
operators $\{\hat U_P, P \in \mathcal P (\Gamma)\}$ generating Poincar\'e 
transformations through the action $\hat U_P : \mathcal H \to \mathcal H$
can then be seen to take the form 
\begin{eqnarray} 
{U(x, \bar y)} = K(L^a_b\,x^b+B^a, {\bar y}^a) ,
\label{eq:unitary}
\end{eqnarray} 
where $K(x, \bar y)$ is the Bergman kernel (\ref{Bergman kernel}). 
To see that the operator ${\hat U}$ thus defined is unitary, it suffices 
to show that ${\hat U}{\hat U}^\dagger=\hat{\mathds 1}$, where on account 
of (\ref{Bergman kernel0}) the identity operator $\hat{\mathds 1}$ here in the
phase-space coordinate representation is given by the kernel function. 
Then by (\ref{eq:unitary}) we 
have 
\begin{eqnarray} 
{\bar U}(y, \bar x) = K(y^a, L^a_b\,{\bar x}^b+B^a), 
\end{eqnarray} 
and it follows from (\ref{reproducing property}) and  \eqref{Lorentz transformation} that  
\begin{eqnarray}
\int \! U(x, {\bar y}) \, 
{\bar U}(y, {\bar z}) \, \rd\mu_y &=& 
\int \!K(Lx+B, {\bar y}) \, 
K(y, L{\bar z}+B) \, \rd\mu_y \nonumber \\
&= &
K(Lx+B, L{\bar z}+B) \nonumber \\
&=& \left(\frac{2^3\cdot 4 !}{ \,\pi^4} \right) 
\frac{1}{[g_{ab}(L^a_c\,x^c-L^a_c\,{\bar z}^c)
(L^b_d\,x^d-L^b_d\,{\bar z}^d)]^4} \nonumber \\ 
&=& K(x,{\bar z}),
\end{eqnarray}
as desired. It is also
apparent that ${\hat U}_{P'} 
{\hat U}_P={\hat U}_{P'P}$ for all $P,P'\in{\mathcal P}(\Gamma)$, so we  
conclude that (\ref{eq:unitary}) gives a unitary representation of the 
Poincar\'e group on the Hilbert space  ${\mathcal H}= {L}^2(\Gamma,{\cal O})$. 

More generally, a representation of the full 15-parameter conformal 
group can also be identified by use of the kernel function. To see this, 
we consider first the four-dimensional subgroup ${\mathcal S}(\Gamma)$ of 
special conformal transformations, given by 
\begin{eqnarray}
S : z^a \mapsto 
\frac{z^a+z^2\lambda^a}{1+2\lambda \!\cdot \!z+\lambda^2 z^2} , 
\label{eq:special} 
\end{eqnarray}
where $\lambda^a$ is a real four-vector,  and we write
$\lambda \! \cdot \! z=g_{ab}\lambda^a z^b$, $\lambda^2=g_{ab}\lambda^a\lambda^b$, 
and $z^2=g_{ab}z^az^b$. The transformation (\ref{eq:special}) is obtained by 
applying an inversion $z^a\mapsto z^a/(g_{bc}z^b z^c)$, then 
shifting the result by 
$\lambda^a$, and finally applying a further inversion. The unitary 
operator ${\hat U}_S$ generating such a transformation on states in 
${\mathcal H}$ is given by 
\begin{eqnarray}
{U}(x, \bar y) = \frac{1}{(1+2\lambda \!\cdot \!x+\lambda^2 x^2)^4} \, 
K\left( \frac{x^a+x^2\lambda^a}{1+2\lambda \!\cdot \!x+\lambda^2 x^2}, 
{\bar y} \right) . 
\label{eq:z44}
\end{eqnarray} 
To see that ${\hat U}_S$ defines a unitary representation of the group 
${\mathcal S}(\Gamma)$ on ${\mathcal H}$, we let $S\in{\mathcal S}
(\Gamma)$ be parameterized by $\lambda^a$ and $S'\in{\mathcal S}
(\Gamma)$ be parameterized by $\mu^a$, and consider the action of 
${\hat U}_{S'}{\hat U}_S$ on a generic state $\psi(x)\in{\mathcal H}$. 
A calculation gives 
\begin{eqnarray}
{\hat U}_{S'}{\hat U}_S \, \psi(x) = \frac{1}{(1+2\nu \!\cdot \!x+ 
\nu^2 x^2)^4} \, \psi \left( \frac{x^a+x^2\nu^a}
{1+2\nu \!\cdot \!x+\nu^2 x^2} \right) , 
\end{eqnarray}  
where $\nu^a = \lambda^a+\mu^a$, and it follows that  
${\hat U}_{S'}\,{\hat U}_S={\hat U}_{S'\,S}$. 
The remaining component of the conformal group is the one-parameter dilatation group $\mathcal D (\Gamma)$, which consists of
transformations of the form $D: z^a\mapsto\Lambda z^a$, where $\Lambda$ 
is a strictly positive real number. It should be apparent that the unitary operator 
generating a dilatation with parameter $\Lambda$ is  
\begin{eqnarray}
{U}(x, \bar y) = \Lambda^4 \, K(\Lambda x^a , {\bar y}^a) . 
\label{dilatation group} 
\end{eqnarray} 
The corresponding action of a dilatation on a state $\psi(z)\in{\mathcal H}$ is thus 
$D : \psi(z)\mapsto\Lambda^4\,\psi(\Lambda z)$, and we see that 
(\ref{dilatation group}) 
defines a unitary representation of the dilatation group on 
${L}^2(\Gamma,{\cal O})$. 

The idea that we exploit in arriving at explicit 
phase-space representations for unitary operators is the fact that on a 
Hilbert space with a reproducing kernel, every operator admits an integral 
representation \cite{Meschkowski}. Thus, for example, the generator of 
the space-time translation $z^a\to z^a+b^a$, where $b^a$ is a real four-vector, is 
given by $b^a {\hat P}_a$, where ${\hat P}_a=\ri\hbar\partial/\partial x^a$ is 
the four-momentum operator; but the four momentum operator ${\hat P}_a$ 
admits the following phase-space representation: 
\begin{eqnarray}
P_a(z,{\bar w}) =  -\frac{\,3\cdot 2^9\, \hbar\, \ri \,}{ \,\pi^4} 
\frac{g_{ab} (z^b-{\bar w}^b)}{[g_{ab}(z^a-{\bar w}^a)(z^b-{\bar w}^b)]^5} .
\end{eqnarray}

\section{Quantum measurements} 

\noindent To make sense of the notion of quantum detection in a relativistic setting we need positive 
operator-valued measures \cite{DL, Davies, Holevo}. In the present context a POVM can be formed by taking a collection $\Phi$ of positive operators 
$\{ \phi_A(y,{\bar z})\}_{A\in{\cal B}}$ on phase space labelled by elements 
of the  
Borel $\sigma$-algebra ${\cal B}(\Gamma)$ generated by the open subsets of 
$\Gamma$. We require $\Phi$ to have the following properties: (a) $\phi_A(y,{\bar z})$ is 
positive for each $A\in{\cal B}$, (b) $\phi_\Gamma(y,{\bar z})=
K(y,{\bar z})$, and (c) for any countable collection of disjoint sets 
$\{A_n\}_{n \in \mathds N}$ in ${\cal B}$ with union 
$A = \cup_{n \in \mathds N} \, A_n$ it holds that 
\begin{eqnarray}
\phi_A(x,{\bar z}) = \sum_n \phi_{A_n}(x,{\bar z})\,. 
\end{eqnarray}
We consider now a measurement operation appropriate for detecting the location of an event in phase space. The POVM is  defined by 
\begin{eqnarray}
\phi_A(x,{\bar z}) = \int_{y\in A} \! K(x,{\bar y})\, K(y,{\bar z})\, 
\rd \mu_y , \quad A\in{\cal B} . 
\label{POVM}
\end{eqnarray}
The recorded  outcome  of such a measurement will be a measurable set $A$ in phase space: 
for instance, the detection of a particle in a certain space-time region, accompanied 
by a four-momentum taking values in a certain range. One can have in 
mind, for example, the detection of a cosmic ray. 
By \eqref{reproducing property} and \eqref{POVM},  the probability that 
the event lies in the set $A\in{\cal B}$ is 
\begin{eqnarray}
{\mathbb P}(A)  = \int \!\phi_A(y,{\bar z}) \, \rho(z,{\bar y})\,  
\rd \mu_y \, \rd \mu_z = \int_{z\in A} \!
\rho(z,{\bar z})\, \rd \mu_z . 
\end{eqnarray}
We see, in particular, in accordance with our earlier discussion, that $\rho(z,{\bar z})$ is the probability density for the outcome,  and hence that the expectation value of any measurable function $F: \Gamma \to \mathds R$ is given by the integral
\begin{eqnarray}
{\mathbb E}[F]  = \int \! \!F(z,{\bar z}) \, \rho(z,{\bar z})\, \rd \mu_z , 
\end{eqnarray}
which is well-defined and finite providing that 
\begin{eqnarray}
 \int \! \! \big | \, F(z,{\bar z}) \, \big | \, \rho(z,{\bar z})\, \rd \mu_z < \infty . 
\end{eqnarray}
Once a measurement has been performed and the outcome recorded, the state of the 
system changes. To model this we require a transformation operator  of the Krauss type \cite{Krauss 1971, Krauss 1983}\,: 
\begin{eqnarray}
T_A(u,v,{\bar x},{\bar y}) = \int\limits_{w\in A} \!\!\frac{K(u,{\bar w})\, 
K(v,{\bar w}) \, K(w,{\bar x})\, K(w,{\bar y})}{K(w,{\bar w})} \,
\rd \mu_w .
\label{transition operator} 
\end{eqnarray} 
One can verify directly that the partial trace of the state
transformation operator generates the POVM.  
That is, we have 
\begin{eqnarray}
\int T_A(x,y,\bar{y},\bar{z})\,\rd \mu_y = \phi_A(x,\bar{z})   
\end{eqnarray} 
for each $A\in{\cal B}$.
Now suppose that the system is initially in the state $\rho_{\rm in}(y,{\bar v})$. 
Then after the measurement we find that
\begin{eqnarray}
\rho_{\rm out}(u,{\bar x}) = \frac{\int T_A(u,v,{\bar x},{\bar y})\, 
\rho_{\rm in}(y,{\bar v}) \, \rd \mu_v \, \rd \mu_y}
{\int T_A(z,v,{\bar z},{\bar y})\, \rho_{\rm in}(y,{\bar v}) \, 
\rd \mu_v \, \rd \mu_y\, \rd \mu_z} , 
\label{out density matrix} 
\end{eqnarray} 
which represents the transformed state that results when the measurement 
determines that the phase-space event lies in the set 
$A\in{\cal B}$. Substituting (\ref{transition operator}) in (\ref{out density matrix}), and 
making use of the reproducing property (\ref{reproducing property}), we 
deduce that 
\begin{eqnarray}
\rho_{\rm out} (u,{\bar x}) = \frac{ \int_{z\in A} \Psi_z(u,\bar{x}) \, \rho_{\rm in}(z,\bar{z})\, 
\rd \mu_z}{\int_{z\in A}  \rho_{\rm in}(z,\bar{z})\, \rd \mu_z} , 
\end{eqnarray} 
where 
\begin{eqnarray}
\Psi_z(u,\bar{x}) = \frac{K(u,{\bar z})\,K(z,{\bar x})}{K(z,{\bar z})}
\label{Bergman density matrix}
\end{eqnarray} 
is the density matrix associated with the normalized wave 
function  
\begin{eqnarray}
\psi_z(u) =  \frac{K(u,{\bar z})}{\, \, [K(z,{\bar z})]^{1/2}}. 
\label{pure state outcome}
\end{eqnarray}
Then in the limit that the recorded outcome shrinks to a phase-space point, 
we find that 
\begin{eqnarray} 
\rho_{\rm out}(u,{\bar x}) = \Psi_z(u,\bar{x}). 
\label{outgoing state}
\end{eqnarray}
In what follows we shall refer to a pure state of the form \eqref{pure state outcome}
 as a \textit{coherent state} with focus $z$. 
We observe, 
in particular, that by virtue of \eqref{Bergman kernel} we have
\begin{eqnarray}
\psi_z(u) = \frac{8\sqrt{3}}{\pi^2} \,  
\frac{[g_{ab}(z^a-{\bar z}^a)(z^b-{\bar z}^b)]^2}
{[g_{ab}(u^a-{\bar z}^a)(u^b-{\bar z}^b)]^4} . 
\end{eqnarray}
We can also refer to a pure density matrix of the form \eqref{Bergman density matrix} as a coherent state, and we note that the family of such density matrices satisfies a completeness relation of the form
\begin{eqnarray}
\int_{z }\Psi_z(x,{\bar y}) \, K(z,{\bar z}) \, \rd \mu_z = K(x, {\bar y}). 
\end{eqnarray}
That such a relation should hold is characteristic of the properties of coherent states \cite{Peremolov 1972, Gazeau 2014} and follows from the fundamental identity \eqref{fundamental identity}, as does the structure of the POVM given by \eqref{POVM}. It is interesting then to note that the coherent states arising in the present context are in one-to-one correspondence with the so-called ``elementary states" that arise in the theory of zero rest mass fields \cite{Penrose 1972}.

The foregoing analysis shows that when the measurement apparatus detects 
that an event has taken place in a region $A$ of phase space, the 
output state will in general be a mixed state, given by the weighted average of the coherent state $\Psi_z(u, \bar x)$ 
over $z \in A$ with respect to the renormalized density 
\begin{eqnarray}
\rho_A(z,\bar{z}) = \frac {\rho_{\rm in}(z,\bar{z})}
{\int_{y\in A}  \rho_{\rm in}(y,\bar{y})\, \rd \mu_y}.
\end{eqnarray} 
If, however, the record 
shows a specific phase-space point $z$ as the result, then the output  density matrix will be the 
coherent state $\Psi_z(u,\bar{x})$ parameterized by $z$. 
At the other extreme, if the measurement is performed but the outcome is not recorded, then the focus is smeared over the whole of the phase space, representing a decoherence effect, and we obtain
\begin{eqnarray}
\rho_{\rm out}(u,{\bar x}) = \int \Psi_z(u,\bar{x}) \, \rho_{\rm in}(z,\bar{z})\, 
\rd \mu_z . 
\label{eq:71} 
\end{eqnarray} 

\section{Properties of coherent states} 

\noindent 
On the matter of the interpretation of the coherent states, we remark that 
the family of coherent states $\{\psi_z(u)\}$ parameterized by $z\in \Gamma$ is Poincar\'e invariant in the sense that under the unitary 
transformation (\ref{eq:unitary}) one has 
\begin{eqnarray}
{\hat U}\psi_z(u) =  \psi_w(u), 
\end{eqnarray} 
where $w^a =L^a_b\,z^b-B^a$. In other words, 
the action of the unitary representation of the 
Poincar\'e transformation on a coherent state 
$\psi_z(u)$ focussed at $z\in \Gamma$ is the coherent state $\psi_w(u)$ focussed at $w$, where $w$ is the 
result of the corresponding inverse Poincar\'e transformation on $z$. 

More generally, one can show that manifold of coherent states is invariant 
under the action of the 15-parameter conformal group. To see this, 
consider first  the dilatation group. From (\ref{dilatation group}) we see that ${\hat U}\psi_z(u)=\psi_w(u)$, where 
$w^a=\Lambda^{-1}z^a$. Thus, the action of the dilatation on a coherent state $\psi_z(u)$ focussed at the phase-space point $z$ is a coherent state $\psi_w(u)$ focussed at $w$, where $w$ is the 
result of the corresponding inverse dilatation on $z$. 

The action of the  
special conformal transformations on a coherent state is a little more 
subtle. Writing ${\hat U}_\lambda$ for the unitary operator (\ref{eq:z44}), 
we find that  
\begin{eqnarray}
{\hat U}_\lambda \psi_z(u) =  \frac{8\sqrt{3}}{\pi^2} \,  
\frac{[(z-{\bar z})\!\cdot\!(z-{\bar z})]^2}
{(1+2\lambda \!\cdot \!u+\lambda^2 u^2)^4} 
\left[\left(\frac{u+u^2\lambda}{1+2\lambda \cdot u+
\lambda^2 u^2}-{\bar z}\right) \!\cdot\! \left( 
\frac{u+u^2\lambda}{1+2\lambda \cdot u+\lambda^2 u^2}
-{\bar z}\right)\right]^{-4}\!.  
\label{conformal transformation of elementary state} 
\end{eqnarray}
A calculation then gives 
\begin{eqnarray}
\left(\frac{u+u^2\lambda}{1+2\lambda \cdot u+
\lambda^2 u^2}-{\bar z}\right) \!\cdot\! \left( 
\frac{u+u^2\lambda}{1+2\lambda \cdot u+\lambda^2 u^2}
-{\bar z}\right) 
= \frac{1-2\lambda 
\!\cdot\! {\bar z} + \lambda^2 {\bar z}^2}{1+2\lambda \!\cdot\! u + 
\lambda^2 u^2} \left[ (u-{\bar w})\!\cdot\!(u-{\bar w})\right] , 
\end{eqnarray} 
where 
\begin{eqnarray}
w^a = \frac{z^a-z^2\lambda^a}{1-2\lambda \!\cdot \!z+\lambda^2 z^2} . 
\end{eqnarray} 
Thus the terms involving $(1+2\lambda \!\cdot \!u+\lambda^2 u^2)$ 
appearing in (\ref{conformal transformation of elementary state}) cancel, and we obtain 
\begin{eqnarray}
{\hat U}_\lambda \psi_z(u) = \frac{8\sqrt{3}}{\pi^2} \,  
\frac{[(z-{\bar z})\!\cdot\!(z-{\bar z})]^2}
{(1-2\lambda \!\cdot \!{\bar z}+\lambda^2 {\bar z}^2)^4} 
\frac{1}{[(u-{\bar w})\!\cdot\!(u-{\bar w})]^4}. 
\label{phase shifted elementary state}
\end{eqnarray} 
In fact, the right side of (\ref{phase shifted elementary state}) is a phase-shifted version of the coherent 
state $\psi_w(u)$ centred at $w$: 
\begin{eqnarray}
{\hat U}_\lambda \psi_z(u) = \re^{{\rm i}\theta} \, \frac{8\sqrt{3}}{\pi^2} \,  
\frac{[(w-{\bar w})\!\cdot\!(w-{\bar w})]^2}
{[(u-{\bar w})\!\cdot\!(u-{\bar w})]^4} , 
\label{with phase factor}
\end{eqnarray} 
where the phase shift $\theta$ is given by 
\begin{eqnarray}
\theta =  \frac{1}{2\ri} \log\left[ \frac{1-2\lambda \!\cdot \! z+
\lambda^2 z^2}
{1-2\lambda \!\cdot \!{\bar z}+\lambda^2 {\bar z}^2} \right] = 
\frac{1}{2\ri} \log\left[ \frac{z^2 {\bar w}^2}{{\bar z}^2 w^2} \right] . 
\end{eqnarray} 
In other words, the result of the action of a special conformal transformation on a 
coherent state $\psi_z(u)$ focussed at $z$ is 
a coherent state $\psi_w(u)$ focussed at 
$w$ with a phase shift $\theta$, where $w$ is the result of the action of the corresponding 
inverse special conformal transformation on $z$. Since the physical state of a 
system is defined up to an overall phase (or, equivalently, the phase 
factor drops out if we consider the action of a conformal transformation
on a pure-state density matrix), we deduce that the manifold of 
coherent states is invariant under the 15-parameter conformal group. 

In calculations, it turns out to be convenient in many situations to work with Fourier transforms. In fact, there are some remarkable identities that turn out to be useful in this connection.  The Fourier transform of an element $\psi \in {L}^2(\Gamma,{\cal O})$ is defined by
\begin{eqnarray} 
\Psi(p^a)=\int_{\Gamma} \! \exp ({{\rm i} p_a \bar{z}^a}) \, \psi(z^a) \,  
\rd \mu_z \, .
\label{Fourier transform}
\end{eqnarray} 
The inverse Fourier transform is then given by 
\begin{eqnarray} 
\psi(z^a) =  \frac {1}{8\pi^5} \int_{V^+} \! \exp ({{-\rm i}p_a {z}^a}) \, 
\big(p_bp^b\big)^2\, \Psi(p^a)\,\rd^4p,
\label{inverse Fourier map}
\end{eqnarray} 
where the integration is over the interior of the forward cone defined by 
\begin{eqnarray}
V^+ = \{ p^a : p_a p^a > 0, p^0 > 0\}.
\end{eqnarray}
The argument for \eqref{inverse Fourier map} can be sketched as 
follows. Let $I(z^a)$ denote the outcome of the integral appearing on the 
right-hand side of \eqref{inverse Fourier map}. If we substitute 
\eqref{Fourier transform} into the formula for $I(z^a)$, we get
\begin{eqnarray} 
I(z^a) = \frac {1}{8\pi^5} \int_{V^+} \! \exp ({{-\rm i}p_a {z}^a}) \, 
\big(p_bp^b\big)^2 \int_{\Gamma}  \exp ({{\rm i} p_a \bar{w}^a}) \, \psi(w^a) \,  
\rd \mu_w \,\rd^4p.
\end{eqnarray} 
Then, reversing the order of integration, we have
\begin{eqnarray} 
I(z^a) = \frac {1}{8\pi^5}  \int_{\Gamma} \, \left [\int_{V^+} \! \exp ({{-\rm i}p_a {z}^a}) \, 
\big(p_bp^b\big)^2   \exp ({{\rm i} p_a \bar{w}^a}) \, \rd^4p \right] \, \psi(w^a) \,  
\rd \mu_w .
\end{eqnarray} 
Now, the inner integral can be carried out explicitly, and we obtain
\begin{eqnarray} 
 \frac {1}{8\pi^5} \int_{V^+} \! \exp ({{-\rm i}p_a {z}^a}) \, 
\big(p_bp^b\big)^2   \exp ({{\rm i} p_a \bar{w}^a}) \, \rd^4p = K(z, \bar w), 
\label{Fourier identity 1}
\end{eqnarray} 
where the Bergman kernel is defined as in \eqref{Bergman kernel}. An application of the reproducing property \eqref{reproducing property} then 
shows that $I(z^a)=\psi(z^a)$, and thus we obtain  
\eqref{inverse Fourier map}, the Fourier inversion formula. 

Alternatively, suppose that we are given a map 
$\Psi: V^+ \to {\mathds C}$ on the positive cone such that 
\begin{eqnarray}
\int_{V^+}  \big(p_bp^b\big)^2 \, \big | \Psi(p^a) \big |^2 \,
\rd^4p < \infty ,
\end{eqnarray} 
and we define a holomorphic function $\psi: \Gamma \to {\mathds C}$ 
by use of  \eqref{inverse Fourier map}. 
Let $J(p^a)$ denote the outcome of the integral appearing on the 
right side of (\ref{Fourier transform}). Then for $J(p^a)$ we obtain 
\begin{eqnarray} 
J(p^a) &=& \frac {1}{8\pi^5} \int_{\Gamma} \! \exp ({{\rm i} p_a \bar{z}^a})  
 \int_{V^+} \! \exp ({{-\rm i}q_a {z}^a}) \, 
\big(q_bq^b\big)^2\, \Psi(q^a)\,\rd^4q \,  \rd \mu_z  \nonumber \\ &=& 
\frac {1}{8\pi^5}  \int_{V^+}  \left [ \, \int_{\Gamma} \! \exp ({{\rm i} p_a \bar{z}^a} {{-\rm i}q_a {z}^a})  
  \, \rd \mu_z \,\right ]
\big(q_bq^b\big)^2 \, \Psi(q^a)\,\rd^4q .
\label{inverse argument}
\end{eqnarray} 
Let us consider the inner integration first. 
Writing $z^a=x^a-\ri r^a$ with $x^a$ real, and with $r^a$ timelike and future pointing, we have  
\begin{eqnarray}
{\ri} p_a \bar{z}^a  -{\ri} q_a {z}^a = {\ri}(p_a-q_a) x^a -(p_a+q_a)r^a . 
\end{eqnarray} 
Since $\rd\mu_u=\rd^4x\,\rd^4r$, we find that the $x$-integration 
over Minkowski space gives a delta function. Thus setting $\xi_a=p_a+q_a$ we have
\begin{eqnarray}
\int_{\Gamma} \! \exp ({{\ri} p_a \bar{z}^a} {{-\rm i}q_a {z}^a})  \, \rd \mu_z 
= (2\pi)^4\, \delta^4(p^a - q^a) \int_{V^+} \exp (- \xi_a r^a) \,\rd^4r  \,.
 \end{eqnarray} 
For the $r$-integration, 
we can pass to spherical coordinates. 
Then if we set 
$\xi^2=(\xi^1)^2+(\xi^2)^2+(\xi^3)^2$ and $R^2=(r^1)^2+(r^2)^2+(r^3)^2$, 
because $r^a$ is time-like we find 
\begin{eqnarray}
\int_{V^+} \exp (- \xi_a r^a) \, \rd^4r 
= \int\limits_{r^0=0}^\infty 
\int\limits_{R=0}^{r^0} \int\limits_{\theta=0}^{\pi} \int\limits_{\phi=0}^{2\pi} 
\exp \left({-\xi_0r^0 + \xi R \cos\theta} \right) \, R^2 \rd r^0  \rd R 
 \sin\theta \, \rd\theta \, \rd\phi\, . 
\end{eqnarray} 
Because $r^a$ is future pointing, 
the integrals can now be performed explicitly to give 
\begin{eqnarray}
\int_{V^+} \exp (- \xi_a r^a) \, \rd^4r = \frac{8\pi}{(\xi_a\xi^a)^2} \, , 
\end{eqnarray} 
from which it follows that
\begin{eqnarray}
\int_{\Gamma} \! \exp ({{\ri} p_a \bar{z}^a} {{-\ri}q_a {z}^a})  
  \, \rd \mu_z =    \frac{8\pi^5}{(q_a q^a)^2}\,  \delta^4(p^a - q^a).
 \end{eqnarray} 
Inserting this expression back into \eqref{inverse argument} we immediately see that the result of the integral is $\Psi(p^a)$. 

By a similar argument it follows from \eqref{Fourier identity 1} that 
if $\Psi(p)$ is the Fourier transform of a square-integrable holomorphic 
function $\psi(z)$, and $\Phi(p)$ is 
the Fourier transform of a square-integrable holomorphic function $\phi(z)$, then we have a Parseval identity of the form 
\begin{eqnarray}
\int_{\Gamma} \, \, \psi(z^a) \,  \bar \phi(\bar z^a) \,
\rd \mu_z = \frac {1}{8\pi^5} \int_{V^+}  
\Psi(p) \, (p_a p^a)^2 \, {\bar \Phi(p)} \, \rd^4 p . 
\end{eqnarray}

In the case of a coherent state $\psi_z(u)$ with focus $z \in \Gamma$, a 
calculation shows that its Fourier transform is given by
\begin{eqnarray} 
\Psi_z(p^a) =  \frac{8\sqrt{3}}{\pi^2} \, 
\big[\,g_{ab}(z^a-{\bar z}^a)(z^b-{\bar z}^b)\,\big]^2\, 
 \exp({{\rm i}\bar{z}_a p^a}) \, . 
\end{eqnarray} 
Now writing $z^a=x^a-\ri r^a$, we have $z^a-{\bar z}^a=-2\ri r^a$, so 
\begin{eqnarray} 
\Psi_z(p^a) =  \frac{2^7\!\sqrt{3}}{\pi^2} \, \big(r_ar^a\big)^2\,  
\re^{ -r_a p^a} \, \re^{{\rm i} x_a p^a} . 
\end{eqnarray} 
This relation shows that as $z^a$ varies the Fourier component $\Psi_z(p^a)$ behaves like a plane wave in Minkowski space that has been extended into the future tube, but is damped exponentially for large $r^a$. We notice, in particular, that when $r^a$ is large, corresponding to the case where focal point lies in a low-mass region of $\Gamma$, the damping of the high-energy Fourier components is significant. 

\section{Phase-space localization} 

\noindent 
With a view to getting a better understanding of the degree of localization in phase space that might be achievable in such a detection experiment, let us 
consider properties of the coherent states in more detail. 

For each choice of the focal point $z \in \Gamma$, the associated coherent state  is represented by the normalized wave function $\psi_z(u)$. Now, if $| \phi \rangle \in L^2(\mathcal H, \mathcal O)$ is any other normalized state,
we have the Cauchy-Schwartz inequality
\begin{eqnarray} 
\langle \, \psi_z \, | \,\phi \,\rangle  \,  \langle \, \phi\, | \, \psi_{z} \, \rangle
\leq 1 .
\label{Cauchy Schwartz}
\end{eqnarray}
It then follows immediately from \eqref{reproducing property}, \eqref{pure state outcome} and \eqref{Cauchy Schwartz} that
\begin{eqnarray} 
\phi(z) \,  \bar\phi(\bar z) \leq K(z, \bar z) 
\label{phase-space inequality} 
\end{eqnarray}
for all $z \in \Gamma$, which shows that the amplitude of any pure state at a phase-space point $z$ is bounded by the amplitude at $z$ of the coherent state with focus $z$. 

Hence the coherent states are the most sharply peaked states, and the peak of a coherent state occurs at its focus. It thus makes sense that if the measurement outcome takes the form of a specific point in phase space, then the transformed state should be peaked as much as possible at that point, and hence a coherent state with that peak, as we obtained in \eqref{outgoing state}. 

An interesting physical interpretation of the inequality \eqref{phase-space inequality} can be deduced if we write 
${\rm Im}\,(z^a) = -r^a$ and make use of the phase-space correspondence \eqref{phase-space correspondence}. It follows from 
\eqref{Bergman kernel} that
\begin{eqnarray} 
K(z,{\bar z}) = \frac{3}{ \,4 \, \pi^4 \,\hbar^8}
\left(g_{ab}\,p^a p^b\right)^4 . 
\label{Bergman kernel with momentum}
\end{eqnarray} 
Then if we let
\begin{eqnarray}
M_z = \left(g_{ab}\,p^a p^b \right)^{1/2}
\end{eqnarray} 
denote the mass associated with the phase-space point $z^a$, we can write \eqref{phase-space inequality} in the form of a localization bound on the probability density. In particular, we obtain 
\begin{eqnarray} 
\rho(z, \bar z) \leq \frac{3}{ \,4\, \pi^4 \,\hbar^8}\, M_z^{\,8} \, ,
\label{mass inequality} 
\end{eqnarray}
which shows that states cannot be localized very sharply in regions of phase space 
with low mass, but that for higher mass a much greater degree of localization 
can be achieved. 

We can thus think of (\ref{mass inequality}) as a \textit{localization theorem for 
relativistic quantum theory}. Suppose that a phase-space event 
of a relativistic system is characterized by a pure state $\phi(z)$. Then 
the probability of detecting the event is determined by the normalized density function 
\begin{eqnarray} 
\rho(z,{\bar z})=\phi(z) \,  \bar\phi(\bar z) \, .
\end{eqnarray}
 The localization theorem shows that the maximum value that the density 
function can take, at any given point in the eight-dimensional phase space, is
\begin{eqnarray} 
\rho_{\rm max}(z, \bar z) =  \frac{3}{ \,4\, \pi^4 \, \lambdabar_z^{\,8} }\, ,
\end{eqnarray}
where $\lambdabar_z$ denotes the reduced 
Compton wavelength  associated with the phase-space point 
$z$. Now, if an 
event is detected to have occurred at a specific phase-space point $z$, then the resulting output wave 
function will be given by the corresponding coherent state, for which the bound in 
(\ref{mass inequality}) is saturated. Since the density function 
$\rho(z,{\bar z})$ has to integrate to unity over the phase space, one is thus able to conclude
that for a system of short Compton wavelength the event will be 
highly localized in phase space. It follows that we can view the coherent states as representing in some sense
the most ``classical" type of state that can be formed over the relativistic phase space. 

To gain further intuition about the nature of localization, let us consider an 
example in which the state of the system is given by a holomorphic wave function $|\,\xi\rangle$ which we take to be normalized, and an experimentalist wishes to determine whether a localized event at a phase space point $z$ can be detected. Since this is a ``yes-no" type of question, a projective measurement is appropriate, and accordingly we consider the projection operator 
\begin{eqnarray}
\hat \Pi_z = |\psi_z \rangle \,  \langle\psi_z | \, ,
\end{eqnarray} 
where $|\psi_z \rangle$ denotes a normalized coherent state focussed at the point $z$. Here as an aid to intuition we
introduce the usual bra-ket notation and we add ``hats" to operators. Then 
$|\,\xi\rangle$ can be split uniquely into a part that is localized at $z$ and a part that is orthogonal to $|\psi_z \rangle$, so
\begin{eqnarray}
|\,\xi\rangle = |\psi_z \rangle \,  \langle\psi_z | \, \xi \rangle \, + (\hat {\mathds 1} - \hat \Pi_z)  |\,\xi\rangle \, .
\end{eqnarray} 
In fact, it is not difficult to show that if a holomorphic function $\theta(u) \in L^2(\mathcal H, \mathcal O)$ is orthogonal to a coherent state $\psi_z(u)$ with focal point $z$, then $\theta(u)$ vanishes at the focal point. Thus, we can say in a meaningful sense that
any state that is orthogonal to a coherent sate with focus $z$ is delocalized away from the focal point. 

If the outcome of the projective measurement is affirmative, then, by the usual L\"uders-type rules for projective measurements \cite{Luders 1951},  the transformed state will be the localized state $\psi_z(u)$. Otherwise, we obtain the delocalized wave function given by the uniquely determined relative state. In particular, if the initial state is given by the holomorphic function $\xi(u)$ then the outcome of a projective measurement based on the projection operator associated with the coherent state $|\psi_z \rangle$ will be affirmative with probability 
\begin{eqnarray}
p = \langle\psi_z | \, \xi \, \rangle \langle \, \xi  | \, \psi_z\rangle = 
\frac{\, \xi(z) \, \bar  \xi (\bar z)} {K(z, \bar z) }.
\end{eqnarray} 

Now suppose that the wave function $|\xi\rangle$ is itself a 
coherent state, centred at the phase-space point $w=x-\ri r$. Then we have 
$|\xi\rangle=|\psi_w\rangle$, and the probability $p$ of obtaining a ``yes'' 
outcome in a projective measurement involving the projection operator 
${\hat\Pi}_z$ with $z=x'-\ri r'$ 
is  given by $|\langle\psi_w|\psi_z\rangle|^2$. A 
calculation shows that 
\begin{eqnarray} 
p
=   \left[\,\frac{(w-{\bar w})\cdot(w-{\bar w}) \,\,
(z-{\bar z})\cdot(z-{\bar z}) } {(w-{\bar z})\cdot(w-{\bar z}) \,\,(z-{\bar w})\cdot(z-{\bar w})} \, \right]^{\,4}, 
\end{eqnarray} 
which can be viewed as a cross ratio between the four points $w^a$, $z^a$, 
$\bar w^a$ and $\bar z^a$. In fact, it is well known that the transition 
probability between two pure states in nonrelativistic quantum mechanics 
admits an interpretation as a cross ratio between points in complex 
projective space \cite{Brody Hughston 2001}. 

But in the present context what 
is  surprising is that the cross ratio involves points in complex 
Minkowski space. It is a straightforward exercise to verify that the cross ratio 
is conformally invariant and hence \textit{a fortiori} Poincar\'e invariant.
\section{Discussion} 

\noindent 
In summary, we have shown that the future tube possesses a phase-space geometry appropriate both for (a) formulation of a  consistent Hamiltonian mechanics 
for relativistic systems, and (b) construction of a quantum theory of space-time events. In particular, the Hilbert space of square-integrable holomorphic  functions on the future tube can be interpreted as the pure state space of relativistic quantum mechanics. The resulting structure is rich enough to allow for the development of
a manifestly covariant theory of measurement for the detection of phase-space events. The theory incorporates a natural transformation rule for the
quantum state after the measurement, a concept that has hitherto been 
lacking in relativistic quantum theory. We are also able to gain some understanding of the extent to which relativistic events can be localized. An upper bound for the phase-space probability density can be determined, which is inversely proportional to the eighth power of the Compton wavelength. The upper bound is achieved at any given point in phase space by the probability density associated with the phase-space coherent state that has its focal point at that point. 

\vspace{0.5cm}
\begin{footnotesize}
\noindent {\bf Acknowledgements}. DCB acknowledges support from the Russian Science Foundation, grant 20-11-20226.   We thank M. Hoban for helpful discussions. 
\end{footnotesize}

\end{document}